\documentclass[aps,prl,reprint,superscriptaddress,nofootinbib,floatfix,preprintnumbers]{revtex4-2}
\usepackage{amsmath,amssymb,amsfonts}
\usepackage{graphicx}
\usepackage{hyperref}
\usepackage{xcolor}
\usepackage{comment}
\usepackage{orcidlink}

\newif\ifshowcomments
 \showcommentstrue   

\ifshowcomments
  \newcommand{\es}[1]{{\color{blue}[ES: #1]}}
  \newcommand{\mb}[1]{{\textcolor{orange}{[MB: #1]}}}
  \newcommand{\kf}[1]{{\textcolor{purple}{[KT: #1]}}}
  \newcommand{\mk}[1]{{\textcolor{violet}{[MK: #1]}}}
  \newcommand{\cn}[1]{{\textcolor{magenta}{[CN: #1]}}}
  \newcommand{\mj}[1]{{\textcolor{cyan}{[MJ: #1]}}}
\else
  \newcommand{\es}[1]{}
  \newcommand{\mb}[1]{}
  \newcommand{\kf}[1]{}
  \newcommand{\mk}[1]{}
  \newcommand{\cn}[1]{}
  \newcommand{\mj}[1]{}
\fi

\newcommand{\lettersection}[1]{\par\addvspace{0.7\baselineskip}\noindent\textbf{\textit{#1.}}\enspace}

\newcommand{\GeV}{\mathrm{GeV}}
\newcommand{\Mpc}{\mathrm{Mpc}}
\hypersetup{colorlinks=true,linkcolor=blue,urlcolor=blue,citecolor=blue}

\begin{document}

\title{Little Red Dots from Primordial Vacuum Bubbles}

\author{Miguel Barroso Varela~\orcidlink{0009-0006-9844-7661}}
\email{miguelbarrosovarela@gmail.com}
\affiliation{Centro de F\'{\i}sica das Universidades do Minho e do Porto, Rua do Campo Alegre s/n,
4169-007 Porto, Portugal}

\author{Katherine Freese~\orcidlink{0000-0001-9725-7395}}
\email{ktfreese@utexas.edu}
\affiliation{Texas Center for Cosmology and Astroparticle Physics, Weinberg Institute for Theoretical
Physics, Department of Physics, The University of Texas at Austin, Austin, TX 78712,
USA}
\affiliation{The Oskar Klein Centre, Department of Astronomy, Stockholm University, 106 91 Stockholm, Sweden}

\author{Matthew Kleban~\orcidlink{0000-0002-1889-2487}}
\email{mk161@nyu.edu}
\affiliation{Center for Cosmology and Particle Physics, New York University, New York, NY}
\affiliation{Center for Computational Astrophysics, New York, NY 10010, USA}

\author{Cameron Norton~\orcidlink{0000-0002-4382-9885}}
\email{cen8858@nyu.edu}
\affiliation{Center for Cosmology and Particle Physics, New York University, New York, NY}

\author{Evangelos I. Sfakianakis~\orcidlink{0000-0002-1525-2433}}
\email{evangelos.sfakianakis@austin.utexas.edu}
\affiliation{Texas Center for Cosmology and Astroparticle Physics, Weinberg Institute for Theoretical
Physics, Department of Physics, The University of Texas at Austin, Austin, TX 78712,
USA}
\affiliation{Department of Physics, Harvard University, Cambridge, MA, 02131, USA}

\begin{abstract} 
Quantum tunneling transverse to the inflaton trajectory can produce
$10^6$--$10^8M_\odot$ primordial black holes at the abundance of JWST's
little red dots, without the enhanced curvature perturbations excluded by
limits on CMB spectral distortions. Rare vacuum bubbles nucleate during a
brief interval of inflation and later form black holes, with adjustable mass and abundance and a narrow mass
spectrum of percent-level width. The leading constraint, from accretion, is satisfied for a population
matching the bright little red dots even if all the emitted radiation
escapes as hard X-rays.
\\

\end{abstract}

\preprint{UT-WI-37-2026}
\preprint{NORDITA-2026-111}

\maketitle

\lettersection{Introduction} 
The James Webb Space Telescope (JWST) has revealed compact, high-redshift sources known as little red dots (LRDs), several of which exhibit broad Balmer emission lines consistent with accreting black holes~\cite{Matthee:2023utn,Kocevski:2025tft}. Standard virial analyses of broad-line LRD samples infer black-hole masses of approximately $10^6$--$10^8M_\odot$, while modeling of the host emission places several sources above the local black-hole--stellar-mass relation~\cite{Kocevski:2025tft}. For Abell 2744-QSO1 at $z=7.04$, a virial estimate of about $4\times10^7M_\odot$~\cite{Furtak:2023ege} is consistent with a subsequent dynamical measurement near $5\times10^7M_\odot$~\cite{Juodzbalis2026Dynamical}.
Primordial black hole (PBH) formation within a factor of a few of the observed masses would substantially reduce the required accretion growth and its sensitivity to uncertain feedback physics. This is particularly relevant for systems such as QSO1, where growth histories must also remain compatible with the low stellar content and metallicity~\cite{Zhang:2025oyl,DeLuca:2025nao}. Cosmological simulations with a massive primordial seed can reproduce several properties of QSO1 with only modest subsequent mass growth~\cite{Zhang:2025oyl}. 

Forming PBHs directly at LRD masses from enhanced, approximately Gaussian curvature perturbations (i.e.~the usual PBH production mechanism from inflation) is strongly constrained by CMB spectral distortions. Because collapse occurs only in the tail of the distribution, a substantial abundance requires a large curvature variance, and the many fluctuations that fail to collapse are erased by Silk damping, injecting energy that distorts the CMB spectrum~\cite{Nakama:2017xvq}.
The $\mu$-distortion era, approximately $5\times10^{4}\lesssim z\lesssim2\times10^{6}$, is primarily sensitive to modes with comoving wavenumber $50\lesssim k\lesssim10^{4}\,\mathrm{Mpc}^{-1}$, a range that includes the scales associated with LRD-mass PBHs in the standard radiation-era mapping~\cite{Nakama:2016kfq,Balaji:2026uey}. The precise mass range constrained depends on the perturbation spectrum and collapse prescription, as well as the gradual boundaries of the distortion window~\cite{Nakama:2016kfq,Nakama:2017xvq}. For approximately Gaussian perturbations, the FIRAS constraint~\cite{Fixsen:1996nj,Bianchini:2022dqh} sharply suppresses the primordial abundance above an initial seed-mass scale of roughly $10^3$--$10^4M_\odot$ in the scenarios considered in Ref.~\cite{DeLuca:2025nao}, well below the black-hole masses inferred for LRDs.

\begin{figure}[t]
\centering
\includegraphics[width=\columnwidth]{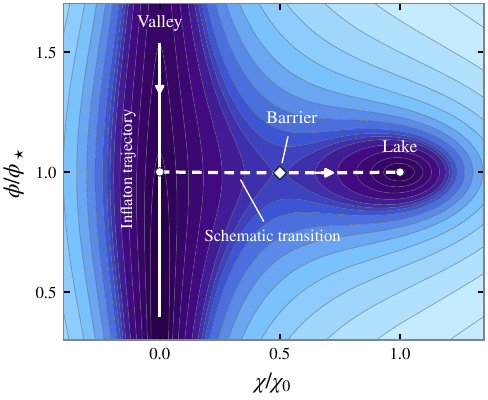}
\caption{Illustrative two-field potential with a rolling valley (the inflaton trajectory) and a nearby basin (``lake") separated by a barrier~\cite{Kleban:2023ugf}. The arrow along the valley indicates the inflaton slow-roll trajectory.  
The dashed line is a schematic for the instanton across the barrier, and $\chi_0$ is the value of the transverse field $\chi$ at the basin minimum.
The tunneling probability across the barrier peaks at $\phi=\phi_\star$. In our scenario, regions of space where the fields tunnel to the ``lake" eventually become PBHs, as explained in the main text.
}
\label{fig:potentialgeometry}
\end{figure}

This has motivated lighter primordial seeds that form on smaller scales and grow by accretion or mergers~\cite{Zhang:2025tgm,DeLuca:2025nao}, and mechanisms that decouple the PBH abundance from the curvature power, such as rare regions with distinct inflationary histories~\cite{Nakama:2016kfq}, strongly non-Gaussian perturbations~\cite{Hooper:2023nnl,Allegrini:2026jqt,Dave:2026gsd,Balaji:2026uey}, or false-vacuum domains surviving a catalyzed transition~\cite{Guo:2026cuv}.
Here we point out that vacuum bubbles nucleated during inflation
can produce supermassive PBHs with no enhancement of the curvature
perturbations, avoiding the spectral-distortion bound on
enhanced curvature
power~\cite{Garriga:2015fdk,Deng:2017uwc,Kleban:2023ugf,HuangPiao:2023}. The
bubbles nucleate by tunneling in a field direction transverse to the inflaton
trajectory (Fig.~\ref{fig:potentialgeometry}), expand to super-horizon size,
and form black holes when they re-enter the horizon after inflation. Each PBH
comes from a single, rare tunneling event, whose probability is set by an
instanton action and not by the amplitude of perturbative fluctuations. The
inflaton perturbations keep their usual slow-roll amplitude near the
transition point, and the transverse field is heavy, with no cosmologically
significant fluctuations of its own apart from the bubbles. The black-hole
mass is exponentially sensitive to the number of $e$-folds remaining when the
bubble nucleated. In earlier work, two of us showed that in a broad class of models nucleation
is naturally localized near a particular value of the inflaton, yielding an
approximately monochromatic PBH spectrum~\cite{Kleban:2023ugf}.
Here we calculate the nucleation rate and the width of the mass spectrum in a controlled approximation, match the abundance to the LRDs, and check the constraints.  While we focus on vacuum bubbles, domain walls or other seeds produced by non-perturbative transitions during inflation could serve as well~\cite{Garriga:2015fdk,Kleban:2023ugf}.

Supermassive PBHs have been proposed as the origin of LRDs~\cite{Huang:2024LRD}, but whether LRDs require primordial seeds is not settled: astrophysical heavy seeds may suffice~\cite{Chon:2026,Natarajan:2023rxq},  supermassive Dark Stars may collapse to black holes~\cite{Freese:2015mta, Ilie:2025zzj, Freese:2025dmo}, 
and the black-hole masses themselves are debated~\cite{Rusakov:2025}.  Regardless, the mechanism of this paper supplies supermassive PBHs without the constraints that exclude the more standard approach.

We set the speed of light and $\hbar$ to unity and normalize the scale factor $a(t)$ to $a_0=1$ today. Here $M_{\rm Pl}=(8\pi G)^{-1/2}$ is the reduced Planck mass. Number densities are comoving, and overdots denote derivatives with respect to cosmic time $t$.

\lettersection{Population targets} Let $n_{\rm LRD}$ be the observed comoving density of a selected LRD sample.  We consider two  LRD number-density targets. One of them is the fiducial number density: the rest-frame UV luminosity functions motivate an integrated-population range $n_{\rm LRD}\sim(2\times10^{-5}$--$10^{-4})\,\Mpc^{-3}$~\cite{Kocevski:2025tft,Akins:2025iqo}. The other is the bright end of the distribution, a useful lower limit for the number density,  $n_{\rm LRD}=10^{-6}\,\Mpc^{-3}$~\cite{Kocevski:2025tft}. The sample selections and bin sums are given in the Supplemental Material. For each target we consider a single PBH population with a narrow mass spectrum that accounts for the corresponding LRD sample; the observed masses follow from the accretion histories considered below.
The comoving number density of this population, $n_{\rm PBH}$, is conserved if mergers are neglected, and we define $q\le 1$ as the fraction of these PBHs that appear in the selected LRD sample,
\begin{equation}
    q\equiv n_{\rm LRD} / n_{\rm PBH}\, .
\end{equation}
Thus $q=1$ gives the minimum PBH number density needed to match the observed LRD density; smaller $q$ requires more PBHs.

Before recombination, Compton drag strongly suppresses infall onto PBHs. Recombination reduces the free electron abundance and weakens this drag, so we neglect the suppressed pre-recombination accretion and begin our accretion histories at $z_i=1100$~\cite{Ricotti:2007jk}.
Let $M_i$ be the PBH mass at $z_i$ and $f_i$ the fraction of dark matter
in the form of PBHs at that time.
 Neglecting growth before $z_i$, $M_i$ is equal to the PBH mass at formation. The  comoving abundance and the fraction required to match the LRD density are
\begin{equation}
n_{\rm PBH}=\frac{f_i\rho_{\rm DM,0}}{M_i}\qquad\text{and} \qquad f_{i, \rm req}=\frac{M_i n_{\rm LRD}}{q\rho_{\rm DM,0}} \, ,
\label{eq:numberdensity}
\end{equation}
Here $\rho_{\rm DM,0} = \rho_{\rm DM}(z_i)/(1+z_i)^3\simeq3.3\times10^{10}M_\odot\Mpc^{-3}$  is the present-day  dark matter density, equal to the comoving dark matter density~\cite{Planck:2018vyg}.
Subsequent accretion is described in terms of the mass-growth factor $\mathcal{G}$ as $M(z)=\mathcal G(z)M_i$, with $\mathcal G(z_i)=1$, so an observed mass must be mapped back to $M_i$ before applying the abundance constraint.

\lettersection{Localized production} 
We describe a narrow PBH mass function by its characteristic formation mass $M_\star$, logarithmic width $\sigma_{\ln M}$ and comoving number density  $n_{\rm PBH}$. In our scenario, all three are determined by how the nucleation rate $\lambda$ (in units of 1/time) of vacuum  bubbles depends on the inflaton.\footnote{We denote by $\Gamma$ the nucleation rate per unit physical volume and time. The corresponding rate in one Hubble volume is $\lambda\equiv\Gamma/H_I^3$, where $H_I$ is the inflationary Hubble scale which is taken as constant.  
}
It reaches its maximum $\lambda_\star$ at a preferred inflaton value $\phi_\star$, which the inflaton crosses $N_\star$ $e$-folds before the end of inflation. Throughout, a subscript $\star$ denotes a quantity evaluated at $\phi = \phi_\star$, or equivalently at $N = N_\star$. For a displacement $\Delta\phi=\phi-\phi_\star$, the rate is~\cite{Kleban:2023ugf} 
\begin{equation}
\lambda(\Delta\phi)\simeq\lambda_{\star}e^{-cR_\star^2(\Delta\phi)^2},
\label{eq:rate}
\end{equation}
where $R_\star$ is the radius of a bubble nucleated at $\phi_\star$ and $c$ is a dimensionless localization coefficient.\footnote{If the inflaton passes more than one such point $\phi_\star$,
bubbles nucleate at several distinct epochs and the mass spectrum has several
narrow peaks, each with its own mass and
abundance~\cite{HuangPiao:2023}. }

The value of $N_\star$ determines the size of these bubbles at collapse, and hence the peak PBH mass [Eq.~\eqref{eq:mass}].
The combination $cR_\star^2$ fixes the duration of production and the width of the peak [Eq.~\eqref{eq:width}]. The peak rate $\lambda_\star$, integrated over that duration, fixes the abundance [Eq.~\eqref{eq:Pn}].  
Nucleation events are independent and produce a population of PBHs that is  Poisson-distributed in space.

A simple realization is a two-scalar field model of the type considered in Ref.~\cite{Kleban:2023ugf}, in which a slow-roll valley (the inflaton trajectory) is separated by a transverse barrier from a lower-energy basin (Fig.~\ref{fig:potentialgeometry}). Most regions remain in the valley, while rare transitions populate the basin. With the techniques of Ref.~\cite{JKMN}, $c$ and the leading semiclassical contribution to the tunneling rate 
can now be calculated in a controlled approximation. We summarize that calculation below; the full derivation and an explicit two-field potential are given in the Supplemental Material.

At approximately constant $H_I$ and $\dot\phi$, 
we can relate the field displacement $\Delta\phi$ to the change in $e$-folding number as $\Delta\phi = -(N-N_\star)\dot\phi/H_I$, where $N$ is the number of e-folds remaining in inflation. Inserting this into Eq.~\eqref{eq:rate} and using
the dimensionless scalar amplitude $\Delta_{\mathcal R}=H_I^2/(2\pi|\dot\phi|)$, we obtain a tunneling rate that is a Gaussian function of the number of $e$-folds 
\begin{equation}
\lambda(N)\propto e^{-(N-N_\star)^2/w_N^2} \, , \quad w_N=\frac{2\pi\Delta_{\mathcal R}}{\sqrt c\,H_IR_\star} \, .
\label{eq:width}
\end{equation}
Here $w_N$ is the width of the nucleation-rate peak in $e$-folds and $N$ counts $e$-folds remaining until the end of inflation. 
For nearly constant $H_I$, the probability of nucleation in an $e$-fold interval $dN$ is approximately
$dP \simeq |dN|{\lambda(N)}/{H_I}\,$.
As seen in Eq.~\eqref{eq:width}, if the  amplitude of the curvature power spectrum during bubble nucleation is taken to be similar to the CMB-scale amplitude $\Delta_{\mathcal R}^2=2.1\times10^{-9}$~\cite{Planck:2018vyg}, the width $ w_N$ is naturally small, but could be increased significantly for smaller $H_IR_\star$ or larger $\Delta_{\mathcal R}$~\cite{Kleban:2023ugf}.

A bubble nucleated $N$ $e$-folds before prompt reheating has radius
   $R_{\rm rh}\simeq H_I^{-1}e^N$ at reheating, and collapses to a black hole
   when it re-enters the horizon during radiation domination~\cite{Garriga:2015fdk,Deng:2017uwc}, so that
\begin{equation}
M_i\sim\frac{e^{2N}}{GH_I} \, .
\label{eq:mass}
\end{equation}
Here $M_i$ is the initial primordial black hole mass.
All masses and widths quoted below are values at formation. The normalization of Eq.~\eqref{eq:mass} carries an order-unity uncertainty from the collapse and depends on the reheating history~\cite{Deng:2017uwc,Kleban:2023ugf}. For the  scale $H_I=10^{10}\GeV$ used below, the interval $10^6$--$10^8M_\odot$ corresponds to $N\simeq40.15$--$42.45$. For the Gaussian profile in Eq.~\eqref{eq:width}, the number-weighted standard deviation\footnote{The characteristic spread $\Delta M/M\simeq2w_N$ differs from this standard deviation by a factor of $\sqrt{2}$.} is $\sigma_{\ln M}=\sqrt2w_N$, including inflationary dilution.

\lettersection{Abundance requirements} Assuming one PBH forms per bubble, the integrated nucleation probability required to match the observed LRD density is, for a monochromatic population,\footnote{The $M^{3/2}$ scaling follows from combining the inflationary dilution $e^{-3N_\star}$ with the mass relation $M_\star\propto e^{2N_\star}$; the normalization assumes prompt reheating and the Standard Model thermal history, with the derivation given in the Supplemental Material.}
\begin{equation}
\mathcal P_{\rm req}\simeq\frac{6.8\times10^{-13}}{q}\left(\frac{n_{\rm LRD}}{10^{-4}\, \Mpc^{-3}}\right)\left(\frac{M_i}{10^8M_\odot}\right)^{3/2} \, .
\label{eq:requiredprob}
\end{equation}
For the narrow spectrum considered here, we evaluate Eq.~\eqref{eq:requiredprob} at the peak initial mass, $M_i\simeq M_\star$.  The nucleation history predicts
\begin{equation}
\mathcal P_n = \int dN\,e^{-3(N-N_\star)}
\frac{\lambda(N)}{H_I} \, .
\label{eq:Pn}
\end{equation}
Here $\lambda(N)/H_I$ is the nucleation probability per $e$-fold, and the exponential accounts for the additional dilution of bubbles nucleated earlier than $N_\star$. We require $\mathcal P_n\simeq\mathcal P_{\rm req}$. For the narrow Gaussian profile of Eq.~\eqref{eq:width}, the dilution factor varies little across the production interval, leading to $\mathcal P_n\simeq\sqrt{\pi}w_N\lambda_{\star}/H_I$.
Writing the physical nucleation rate per unit volume and time as $\Gamma=uR^{-4}e^{-B}$,  with $R$ the bubble radius and using $\lambda=\Gamma/H_I^3$, gives
\begin{equation}
\frac{\lambda_\star}{H_I}=\frac{\Gamma_\star}{H_I^4}
\simeq\frac{u e^{-B_\star}}{(H_IR_\star)^4} \, ,
\label{eq:peak}
\end{equation}
where $B$ is the dimensionless Euclidean action difference and $u$ is the dimensionless fluctuation prefactor~\cite{Coleman:1977py,Callan:1977pt}. The abundance therefore constrains the action together with the bubble radius and duration of production.

\lettersection{Rate from the rolling background} Let $\phi_f$ denote the value of the inflaton on the slow-roll trajectory at the time a bubble nucleates, so that far from the bubble the fields are $(\phi,\chi)=(\phi_f,0)$, with $\chi$ the transverse field. 
Since the inflaton rolls, $(\phi_f,0)$ is not a stationary point of the potential $V(\phi,\chi)$ and cannot serve directly as the boundary value of a Euclidean bounce. For any homogeneous background $b(\tau)$, where $\tau$ is Euclidean time, that obeys its equation of motion, however, the action of a localized fluctuation $\xi$ about it is  $S[b+\xi]-S[b]=\int d^4x\,[\tfrac12(\partial\xi)^2+V_b(\xi)]$~\cite{Yamada:2026xsd,JKMN}, with\footnote{This expression is exact in flat space, with cosmological corrections that scale as powers of $H_I R$.}
\begin{equation}
V_b(\xi)=V(b+\xi)-V(b)-\nabla V(b)\cdot\xi \, .
\label{eq:frozen}
\end{equation}
When the background changes little over the size of the bubble, it can be  approximated as frozen at its value at nucleation, $b=(\phi_f,0)$. Equation~\eqref{eq:frozen} then defines an ordinary tunneling problem with a stationary point at $\xi=0$: the slope of the valley cancels identically, its curvature survives only as a negligible mass for the inflaton fluctuation, and the dependence on $\phi_f$ arises entirely from the coupling between the two fields, which localizes the transition. 
The action $B_\star$ entering the peak rate in Eq.~\eqref{eq:peak} is the Coleman action of $V_b$ at $\phi_f = \phi_\star$. 
Expanding the action to second order in $\Delta\phi=\phi_f-\phi_\star$  gives
\begin{equation}
B(\Delta\phi)\simeq B_\star+2\pi^2R_c^2(\Delta\phi)^2 \, ,\quad c=2\pi^2\left(\frac{R_c}{R_\star}\right)^2 \, ,
\label{eq:cmain}
\end{equation}
which leads directly to Eq.~\eqref{eq:rate}. Here $R_c$ is defined by the response of the inflaton to a change in its
value far from the bubble: outside the bubble this response is a harmonic
function in four Euclidean dimensions, $\propto1-R_c^2/r^2$, where $r$ is the Euclidean radial coordinate.  Freezing the background is accurate when the inflaton barely moves while a
bubble forms. A bubble forms in a time of order $R_\star$, a fraction $H_IR_\star$ of an
$e$-fold, while production lasts $w_N$ $e$-folds, so the expansion parameter is
$\alpha\equiv H_IR_\star/w_N\simeq0.084$ for the benchmark below. At first order in $\alpha$ the time dependence enters the Euclidean exponent only through a term odd in Euclidean time, which integrates to zero on the time-symmetric bounce. The leading correction is second order, $\delta B\simeq-0.57\,\alpha^2\simeq-4\times10^{-3}$. The derivation, error budget, and numerical procedure are given in the Supplemental Material.

\begin{figure}[t]
\centering
\includegraphics[width=\columnwidth]{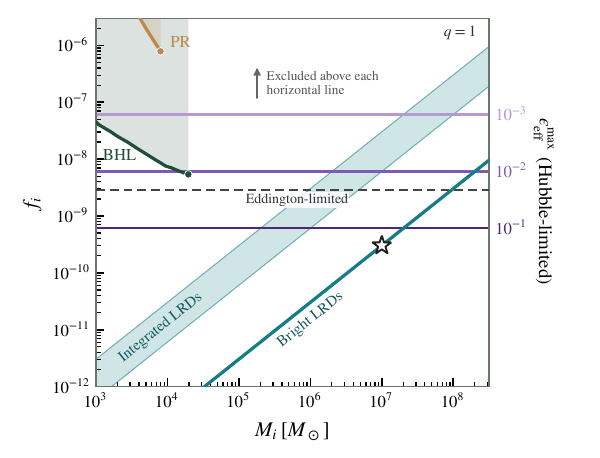}
\caption{
Initial PBH mass fraction $f_i$ versus initial mass $M_i$ at accretion onset, $z_i=1100$. The teal band and line show the abundances required by Eq.~\eqref{eq:numberdensity} for the integrated LRD population, $n_{\rm LRD}=(2\times10^{-5}$--$10^{-4})\,\Mpc^{-3}$, and the bright subset, $n_{\rm LRD}=10^{-6}\,\Mpc^{-3}$~\cite{Kocevski:2025tft,Akins:2025iqo}, both for $q=1$ (every PBH appears as an LRD).  For $q<1$ both shift upward by $1/q$.  Horizontal lines are our estimated CMB upper bounds on $f_i$ from Eq.~\eqref{eq:accretion_limits}: solid for Hubble-limited accretion with $\epsilon_{\rm eff}=0.1$, $10^{-2}$, and $10^{-3}$ from bottom to top, and dashed for Eddington-limited accretion with $\epsilon=0.1$ and $f_X=1$. The right axis gives the largest $\epsilon_{\rm eff}$ allowed at each $f_i$ for Hubble-limited accretion. The green Bondi--Hoyle--Lyttleton (BHL)~\cite{Serpico:2020ehh} and orange Park--Ricotti (PR)~\cite{Park:2010yh,Park:2011rf,Park:2012cr} curves are the published accretion bounds compiled in Ref.~\cite{DeLuca:2025nao}, shaded over their stated mass ranges. The star corresponds to the benchmark point shown in the middle row of Table~\ref{tab:targets}.}
\label{fig:accretion}
\end{figure}

As an example, for the explicit potential specified in the Supplemental Material [Eq.~\eqref{eq:benchmarkpotential}], 
with $H_I=10^{10}\GeV$, the same reference curvature power $\Delta_\mathcal{R}^2$, and $u=1$, matching the scales and epoch to the bright sample with $n_{\rm LRD}=10^{-6}\, \Mpc^{-3}$ and $M_\star=10^7M_\odot$ gives
\begin{equation}
B_\star\simeq57.2\, ,~ H_IR_\star\simeq2.41\times10^{-3} \, ,~
\sigma_{\ln M}\simeq0.0404 \, .
\label{eq:localbenchmark}
\end{equation}
Here $R_\star$ is the Euclidean radius at which the transverse field has fallen halfway from its value at the bubble center to its value far from the bubble, $\chi=0$, i.e. $\chi(R_\star)=\chi(0)/2$. We have not calculated the prefactor, which would shift $B_\star$ by a few. 
Table~\ref{tab:targets} illustrates three cases in which the number density of PBHs matches the bright LRD abundance $n_{\rm LRD}=10^{-6}\, \Mpc^{-3}$ with $q=1$, for three masses  $M_\star/M_\odot=\{10^6, 10^7, 10^8\}$. Adjusting the potential normalization, field scale, and
production epoch for each target mass, while keeping
the dimensionless potential shape fixed,
we find that their mass widths are  close to $4\%$.
The $M_\star=10^7M_\odot$, $n_{\rm LRD}=10^{-6}\,\Mpc^{-3}$, $q=1$ solution is our bright-sample benchmark, marked by the star in Fig.~\ref{fig:accretion}.

For the full-sample reference $n_{\rm LRD}=10^{-4}\, \Mpc^{-3}$, the corresponding actions range from $56.1$ to $49.3$ as the mass increases, while $\sigma_{\ln M}\approx0.04$.  Each matching defines a separate narrow population, and the small rates make bubble overlap negligible. The Supplemental Material gives the scale matching and its mass dependence.

\begin{table}[t]
\caption{Abundance-matched parameter sets for $n_{\rm LRD}=10^{-6}\, \Mpc^{-3}$ and $q=1$, with $H_I=10^{10}\GeV$, $u=1$, and the dimensionless potential parameters of Eq.~\eqref{eq:benchmarkpotential}. The masses listed are PBH masses at formation. Each row denotes a separate choice of dimensional scales and production epoch. 
Actions and widths are obtained by matching the numerical bounce solutions to the target abundance set by JWST LRD data. 
}
\label{tab:targets}
\begin{ruledtabular}
\begin{tabular}{ccccc}
$M_\star/M_\odot$ & $N_\star$ & $B_\star$ & $10^3H_IR_\star$ & $\sigma_{\ln M}$\\
\hline
$10^6$ & $40.15$ & $60.6$ & $2.44$ & $0.0401$\\
$10^7$ & $41.30$ & $57.2$ & $2.41$ & $0.0404$\\
$10^8$ & $42.45$ & $53.8$ & $2.39$ & $0.0408$\\
\end{tabular}
\end{ruledtabular}
\end{table}

\lettersection{Spectral distortion and other bounds} 
The PBH formation mechanism of this paper does not produce significant CMB spectral distortions. 
The enhanced Gaussian curvature fluctuations of the standard PBH formation mechanism source acoustic waves whose dissipation by Silk damping  over the approximate redshift interval $5\times10^4<z<2\times10^6$ produces a dimensionless chemical-potential distortion $\mu$~\cite{Nakama:2017xvq,Chluba:2011hw}. Defining $\overline{\mu}$ as the sky-averaged distortion, a reanalysis of the COBE/FIRAS data gives $|\overline{\mu}|<4.7\times10^{-5}$ at 95\% confidence~\cite{Bianchini:2022dqh}, constraining the curvature variance on the scales relevant for Gaussian PBH formation. The tunneling mechanism of this paper, on the other hand, sets the abundance through a rare transition and does not require  enhanced curvature variance.

Bubble walls can nevertheless launch shocks and compensating underdensity waves. The idealized reflecting-wall estimate of Ref.~\cite{Deng:2018cxb} gives
\begin{equation}
\overline\mu\sim\frac{6.4\times10^{-11}}{q}\left(\frac{n_{\rm LRD}}{10^{-4}\, \Mpc^{-3}}\right)\left(\frac{M_i}{10^8M_\odot}\right)^2.
\label{eq:scaledmu}
\end{equation}
The mass entering this formation-related estimate is the primordial PBH mass, before accretion begins. At $q=1$, this estimate lies well below the FIRAS limit for both population targets. 
Equation~\eqref{eq:scaledmu} estimates the sky-averaged $\mu$ distortion, bounded by the FIRAS monopole limit; the Planck $\mu$-map analysis instead constrains the statistical occurrence of rare, localized patches with unusually large $\mu$.
Under Poisson spatial statistics, the largest expected patch at the full-sample, upper-mass reference has $\mu_{\max}\sim3.2\times10^{-9}$, far below the approximate rare-patch limit $\mu\lesssim10^{-4}$ inferred by Deng et al.~\cite{Deng:2018cxb} from the Planck $\mu$-map distribution of Ref.~\cite{Khatri:2015tla}. The acoustic and occupancy assumptions are detailed in the Supplemental Material. 
For the narrow spectrum considered here, replacing $M^2$ in Eq.~\eqref{eq:scaledmu} by the number-weighted moment $\langle M^2\rangle$ gives a correction of only $\sim0.2\%$ within the Gaussian approximation. 

Because the PBHs are discrete and Poisson distributed, they produce
an irreducible white-noise PBH isocurvature component even without an enhanced Gaussian curvature spectrum \cite{Afshordi:2003zb,Inman:2019wvr,Gerlach:2026}. Applying the CMB+BAO limits of Ref.~\cite{Gerlach:2026} to the white-noise
spectrum $\mathcal P_{\rm iso}=f_i^2k^3/(2\pi^2n_{\rm{PBH}})$, we find an amplitude that is at least
four orders of magnitude below current sensitivity for all masses and abundances considered here at $q=1$.
Since the amplitude scales as $1/q$, our model remains compatible with these limits for $q\gtrsim 10^{-4}$.
Therefore, our model safely evades bounds from $\mu$-distortions and isocurvature.

Random nucleation also implies that the PBHs cluster like matter on large scales, with bias $b_{\rm PBH}\simeq1$~\cite{Hutsi:2026,Wang:2026}.  
Improved clustering measurements could distinguish this
scenario from models in which LRDs preferentially occupy strongly biased
massive halos.

\lettersection{Accretion}
The most important constraint that does not depend on how the PBHs formed comes from accretion: radiation from gas falling onto the black holes after recombination ionizes the intergalactic gas and alters the CMB temperature and polarization spectra~\cite{Ali-Haimoud:2016mbv,Poulin:2017bwe}. Published bounds assume quasi-steady inflow, which fails above about $10^4M_\odot$ where the inflow time approaches the expansion time~\cite{Ricotti:2007jk,Serpico:2020ehh}. The accretion physics at higher masses is uncertain~\cite{Park:2010yh,Park:2012cr,Sugimura:2020rdw,Facchinetti:2022kbg,Agius:2024ecw,Piga:2022ysp}. We do not attempt a systematic treatment here. Instead we ask how aggressive the assumptions must be for the constraint to matter, using two histories that begin at $z_i$ and neglect feedback. In \emph{Hubble-limited accretion} baryons are supplied at the rate $\dot M_{\rm in}=\tfrac32(\rho_b/\rho_{\rm tot})MH$, the mean baryon flux through the radius at which the gravitational time equals the Hubble time $H^{-1}$,\footnote{We thank Yacine Ali-Ha\"{\i}moud for discussions on this point. See the Supplemental Material for the derivation details.} and we assume that all of it reaches the black hole. The luminosity $L=\epsilon\dot M_{\rm in}$ may then exceed the Eddington luminosity $L_E$; for radiative efficiency $\epsilon=0.1$, $L/L_E\simeq13$ at $z_i$. In \emph{Eddington-limited accretion} the inflow is reduced so that $L\leq L_E$. The CMB responds to the part of the luminosity that escapes the source as hard X-rays. We write that part as $\epsilon_{\rm eff}\dot M_{\rm in}$ with $\epsilon_{\rm eff}=\epsilon f_X$, where $f_X\leq1$ is the fraction of the luminosity emitted as an escaping hard continuum. Taking $f_X=1$ gives the strongest constraint. Normalizing our estimate of the CMB response to the Planck-based limit of Ref.~\cite{Serpico:2020ehh} for continuous Eddington emission gives (Supplemental Material)
\begin{equation}
f_{i,\max}\simeq
\begin{cases}
\dfrac{6.1\times10^{-11}}{ \epsilon_{\rm eff}} \,, & \text{Hubble-limited},\\[6pt]
\dfrac{2.9\times10^{-9}}{f_X} \, , & \text{Eddington-limited, }\epsilon=0.1.
\end{cases}
\label{eq:accretion_limits}
\end{equation}
These are estimates that depend on the assumed history and spectrum, not a new likelihood analysis.

Figure~\ref{fig:accretion} compares these limits with the abundances required by Eq.~\eqref{eq:numberdensity}. For $M_i=10^7M_\odot$ and $q=1$, the bright LRDs are allowed under either history even if the entire luminosity emerges as hard X-rays. 
The full population is allowed under Eddington-limited accretion if $f_X\lesssim0.1$--$0.5$, and under Hubble-limited accretion if $\epsilon_{\rm eff}\lesssim0.2$--$1\%$. These requirements tighten in proportion to $M_i/q$, where a mass $M$ observed at redshift $z$ corresponds to $M_i=M/\mathcal G(z)$, with growth factor $\mathcal G\simeq2.3$--$2.8$ at $z=5$. For comparison, observed active nuclei emit of order $10\%$ of their luminosity in hard X-rays~\cite{Duras:2020} and LRDs are X-ray weak~\cite{Yue:2024,Maiolino:2025}, although the relevance of either to isolated black holes before reionization is uncertain.

\lettersection{Discussion} 
Rare vacuum bubbles nucleated during slow-roll inflation 
can produce PBHs  with the mass and abundance needed to explain LRDs. We computed the tunneling rate 
in a field direction transverse to that of the slow-roll motion of the inflaton using a controlled adiabatic approximation.  An explicit two-field realization, shown in  Fig.~\ref{fig:potentialgeometry}, leads to a narrow PBH mass spectrum with logarithmic width $\sigma_{\ln M}\simeq0.04$. The peak PBH mass is set by the time during inflation where tunneling is most effective, while the  abundance is set by the magnitude of the peak tunneling rate; thus they can be independently varied. 
Because this mechanism does not rely on enhanced curvature perturbations, it is not subject to the spectral-distortion bounds on such enhancements.
Distortions sourced by the bubbles themselves are negligible at the required abundance, as are formation-related Poisson  isocurvature perturbations. 
An alternative realization of the orthogonal tunneling mechanism within the chain inflation framework~\cite{Freese:2004vs, Cline:2011fi}, where the universe tunnels through a series of ever lower energy vacua, will be presented separately~\cite{upcoming}.

The most important remaining constraint, from accretion after recombination, is satisfied for the bright LRDs under conservative assumptions. For the full population at $M_i=10^7M_\odot$ it requires that less than $0.2$--$1\%$ of the accreted rest-mass energy escape as hard X-rays for Hubble-limited accretion, or less than $10$--$50\%$ of the luminosity for Eddington-limited accretion.

In summary, we have proposed a mechanism that produces supermassive PBHs by
tunneling transverse to the inflationary trajectory. These PBHs could
explain the little red dots discovered by JWST without violating CMB
spectral-distortion bounds, and are consistent with accretion bounds for
the bright LRD population.

\medskip

\lettersection{Acknowledgments} We thank Yacine Ali-Ha\"{\i}moud, Matthew C.~Johnson, and Anthony Taylor for useful discussions. KF and EIS are grateful for support from the Jeff \& Gail Kodosky Endowed Chair in Physics at the University of Texas.  KF and EIS acknowledge support by the U.S. Department of Energy, Office of Science, Office of High Energy Physics program under Award Number DE-SC0026683.
The work of MBV is supported by FCT (Funda\c{c}\~{a}o para a Ci\^{e}ncia e Tecnologia, Portugal) through the grant 2024.00457.BD with DOI identifier https://doi.org/10.54499/2024.00457.BD. Centro de F\'{\i}sica do Porto is partially funded by Funda\c{c}\~{a}o para a Ci\^{e}ncia e Tecnologia (FCT) under the grant UID04650-FCUP. MK and CN are supported by NSF grant PHY-2412899.

\bibliographystyle{apsrev4-2}
\bibliography{References}

\clearpage
\onecolumngrid
\section*{Supplemental Material}
\setcounter{section}{0}
\renewcommand{\thesection}{\Alph{section}}
\setcounter{equation}{0}
\renewcommand{\theequation}{S\arabic{equation}}
\renewcommand{\theHequation}{S\arabic{equation}}
\setcounter{figure}{0}
\renewcommand{\thefigure}{S\arabic{figure}}
\renewcommand{\theHfigure}{S\arabic{figure}}
\setcounter{table}{0}
\renewcommand{\thetable}{S\arabic{table}}
\renewcommand{\theHtable}{S\arabic{table}}

\section{Population selections}
For a specified sample and redshift interval, the comoving LRD density follows from the rest-frame UV luminosity function $\Phi_{\rm UV}=dn_{\rm LRD}/dM_{\rm UV}$, where $M_{\rm UV}$ is the absolute AB magnitude at $1450\,\text{\AA}$~\cite{Kocevski:2025tft},
\begin{equation}
n_{\rm LRD}=\int_{\rm selection}\Phi_{\rm UV}(M_{\rm UV},z)\,dM_{\rm UV}\simeq\sum_i\Phi_i\Delta M_{{\rm UV},i}.
\label{eq:luminosityfunction}
\end{equation}
Here $\Phi_i$ is the bin-averaged luminosity function and $\Delta M_{{\rm UV},i}$ is the bin width. Summing the one-magnitude bins in Table 4 of Ref.~\cite{Kocevski:2025tft} gives approximately $8.5\times10^{-5}\, \Mpc^{-3}$ at $4.5<z<6.5$ and $7.5\times10^{-5}\, \Mpc^{-3}$ at $6.5<z<8.5$, with bin centers extending from $-17$ to $-21$ and $-22$, respectively. Together with the differently selected COSMOS-Web density of $2\times10^{-5}\, \Mpc^{-3}$~\cite{Akins:2025iqo}, these sums motivate the integrated-population range used in the paper. The bright-subset reference $n_{\rm LRD}=10^{-6}\, \Mpc^{-3}$ is motivated by the $M_{\rm UV}=-21$ bins, whose densities are close to this value in both redshift intervals~\cite{Kocevski:2025tft}. These selections define alternative abundance targets without imposing a magnitude-to-mass relation or identifying the full observed mass range with one primordial peak.

\section{Dilution and width conventions}
At constant $H_I$, define the raw nucleation probability and its dilution-weighted counterpart by integrating over the production window,
\begin{equation}
\mathcal P\equiv\int dN\,\frac{\Gamma}{H_I^4},\qquad \mathcal P_n\equiv\int dN\,e^{-3(N-N_\star)}\frac{\Gamma}{H_I^4}.
\label{eq:productionintegrals}
\end{equation}
For prompt reheating followed by entropy-conserving cooling, the reheating temperature and present comoving bubble density then satisfy~\cite{Kleban:2023ugf}
\begin{equation}
\begin{aligned}
T_{\rm rh}&=\left(\frac{90}{\pi^2g_{*,\rm rh}}\right)^{1/4}\sqrt{H_IM_{\rm Pl}},\\
n_{\rm PBH}&\simeq\mathcal P_n\,H_I^3e^{-3N_\star}\left(\frac{T_0}{T_{\rm rh}}\right)^3\frac{g_{s,0}}{g_{s,\rm rh}}.
\end{aligned}
\label{eq:dilution}
\end{equation}
Here $T_0\simeq2.35\times10^{-13}\GeV$ is the present photon temperature~\cite{Fixsen:2009ug}. The quantities $g_*$ and $g_s$ count the effective relativistic degrees of freedom in energy and entropy, with subscripts $\rm rh$ and $0$ denoting reheating and today. We use the Standard Model values $g_{*,\rm rh}=g_{s,\rm rh}=106.75$ and $g_{s,0}\simeq3.91$. Combining Eq.~\eqref{eq:dilution} with $e^{2N_\star}\sim GH_IM_\star$ gives $\mathcal P_n\propto n_{\rm PBH}M_\star^{3/2}H_I^{-3/2}T_{\rm rh}^3\propto n_{\rm PBH}M_\star^{3/2}$, which explains the cancellation of $H_I$ in Eq.~\eqref{eq:requiredprob}. The same mass relation places the central benchmark at $N_\star\simeq41.3$.

The Gaussian approximation in Eq.~\eqref{eq:width} gives
\begin{equation}
\mathcal P=\frac{\lambda_{\star}}{H_I}\sqrt{\pi}w_N,\qquad B_\star\simeq\ln\left[\frac{u\sqrt{\pi}w_N}{(H_IR_\star)^4\mathcal P_{\rm req}}\right].
\label{eq:integratedrate}
\end{equation}
The first expression is the raw probability $\mathcal P$. The action estimate neglects the difference between $\mathcal P_n=\mathcal P_{\rm req}$ and $\mathcal P$ in the narrow-window limit. Number dilution introduces a factor $e^{-3(N-N_\star)}$, shifting the mean of the Gaussian without changing its variance. Since $\ln(M/M_\star)=2(N-N_\star)$, its moments, with averages weighted by $dn_{\rm PBH}/d\ln M$, are
\begin{equation}
\left\langle\ln\frac{M}{M_\star}\right\rangle=-3w_N^2,\qquad \sigma_{\ln M}=\sqrt2w_N.
\label{eq:MassDistributionMoments}
\end{equation}
The characteristic interval $2w_N$ quoted for localization is therefore distinct from the standard deviation. The relation $\mathcal P_n/\mathcal P=e^{9w_N^2/4}$ provides a check on the numerical calculation, which retains dilution explicitly.

\section{Local numerical production benchmark}
\label{sec:localbenchmark}
The results of the paper rely on Eq.~\eqref{eq:rate}, not on a particular Lagrangian. To calculate $c$ and $\lambda_{\star}$ explicitly we consider a canonical inflaton $\phi$ and transverse field $\chi$, with potential $V=V_0+gv^4U$. The field scale $v$ defines the dimensionless coordinates $x=(\phi-\phi_\star)/v$ and $y=\chi/v$, in which\footnote{For $\varepsilon=0$, this potential reproduces the form of Ref.~\cite{Kleban:2023ugf} after shifting and rescaling the fields. We introduce the transverse bias to adjust the basin depth while keeping the potential along $y=0$ unchanged, which simplifies constructing an abundance-matched benchmark with a shallow rolling valley.}
\begin{equation}
U=\frac14y^2(y-1)^2-\varepsilon y^2(3-2y)+\frac{\kappa}{2}x^2y^2+s x+\frac{\eta}{2}x^2.
\label{eq:benchmarkpotential}
\end{equation}
Here $V_0$ is the valley energy at $\phi_\star$, while the dimensionless coupling $g$ sets the normalization of $U$. The bias $\varepsilon$ lowers the transverse basin while preserving a barrier, and $\kappa$ couples the two fields. Along the valley $y=0$, $s$ and $\eta$ set the slope and curvature of $U$. We choose $(\varepsilon,\kappa,s,\eta)=(0.060,1,10^{-6},10^{-10})$ as local effective-potential parameters for the numerical benchmark, and write $x_f=(\phi_f-\phi_\star)/v$ for the dimensionless inflaton value at nucleation. At $(x,y)=(0,0)$, the transverse valley curvature is $\partial^2U/\partial y^2=1/2-6\varepsilon$, so the chosen bias preserves the barrier while allowing a lower basin.\footnote{This potential is  fine-tuned and suffers from a significant $\eta$ problem, as is often the case in scalar field inflation models. 
}  

\subsection{Bounce about the rolling background}
Because the inflaton rolls along $y=0$, the far-field value $(x_f,0)$ is not a stationary point of $U$, and a Euclidean solution asymptoting to it would run off along the slope. The problem is instead well-posed for fluctuations about the rolling background. Let $b(\tau)$ be the continuation to Euclidean time of the homogeneous rolling background, which solves its own equation of motion, and write $\boldsymbol X=b+\xi$ with $\xi$ localized in space and Euclidean time. Integrating the cross term $\int\partial_\tau b\cdot\partial_\tau\xi$ by parts, the equation of motion for $b$ cancels the term linear in $\xi$, and the boundary terms at early and late Euclidean time cancel for a bounce that is even in $\tau$. In flat space, where the background obeys $d^2b/d\tau^2=\nabla V(b)$,
\begin{equation}
S[b+\xi]-S[b]=\int d^4x\left[\tfrac12(\partial\xi)^2+V(b+\xi)-V(b)-\nabla V(b)\cdot\xi\right].
\label{eq:frozen-exact}
\end{equation}
This is the construction of Ref.~\cite{JKMN} (where the fluctuation is denoted $\eta$), there applied to a background oscillating about a metastable minimum; here the background rolls slowly along a flat direction, and the same identity holds. For the slow-roll background the equation of motion includes Hubble friction, and the cancellation of the linear term holds on the expanding background, with corrections to the flat-space form of order $(H_IR)^2$. The frozen-background approximation replaces $b(\tau)$ by its value at nucleation, $b=(x_f,0)$. In the dimensionless variables of Eq.~\eqref{eq:benchmarkpotential}, the fluctuation potential of Eq.~\eqref{eq:frozen} is $V_b=gv^4U_b$ with
\begin{equation}
U_b(\xi_x,\xi_y)=\tfrac14\xi_y^2(\xi_y-1)^2-\varepsilon\,\xi_y^2(3-2\xi_y)+\tfrac{\kappa}{2}(x_f+\xi_x)^2\xi_y^2+\tfrac{\eta}{2}\xi_x^2 ,
\label{eq:Ub}
\end{equation}
which does not depend on the slope $s$. It is stationary at $\xi=0$ for every $x_f$, with Hessian eigenvalues $\hat m_x^2=\eta$ and $\hat m_y^2=\tfrac12-6\varepsilon+\kappa x_f^2$ and no mixing, so the far-field point is metastable and the $x_f$ dependence of the action arises entirely from the coupling $\kappa$.

For physical Euclidean radius $r$, introduce $\hat r=\sqrt{g}vr$. The bounce equations and dimensionless action $\widehat B$ are~\cite{Coleman:1977py}
\begin{align}
\boldsymbol\xi''+\frac{3}{\hat r}\boldsymbol\xi'&=\nabla U_b(\boldsymbol\xi),\qquad \boldsymbol\xi'(0)=0,\qquad \boldsymbol\xi(\infty)=0,\\
\widehat B&=2\pi^2\int_0^\infty d\hat r\,\hat r^3\left[\tfrac12|\boldsymbol\xi'|^2+U_b(\boldsymbol\xi)\right],\qquad B=\widehat B/g .
\end{align}
Primes denote derivatives with respect to $\hat r$, and $\nabla$ acts on $(\xi_x,\xi_y)$. CosmoTransitions supplies a path-deformation seed~\cite{Wainwright:2011kj}, which is refined by collocation using analytic derivatives. At large $\hat r$ the solution is matched to the linear tails $\propto K_1(\hat m_a\hat r)/\hat r$ in the two Hessian eigendirections, where $K_1$ is the modified Bessel function of the second kind. The $\xi_x$ tail has $\hat m_x=\sqrt\eta=10^{-5}$ and is therefore $\propto\hat r^{-2}$ over the whole computational domain.  Its contribution to the action converges and is included. We define $\widehat R$ by $\xi_y(\widehat R)=\xi_y(0)/2$, giving $R=\widehat R/(\sqrt{g}v)$. For the reference bounce at $x_f=0$, $\widehat B\simeq61.8$ and $\widehat R\simeq5.33$. This convention applies to the thick-wall profiles obtained here, whose central field values do not reach the lower-energy minimum.

\subsection{Localization coefficient}
\label{sec:localization}
The Gaussian dependence of the rate on the inflaton displacement, Eq.~\eqref{eq:rate}, follows from Eq.~\eqref{eq:Ub} by expanding the action in $x_f$. At $x_f=0$ the potential is even in $\xi_x$, so the bounce has $\xi_x=0$. Since the action is stationary with respect to the fields, its first derivative is the explicit one,
\begin{equation}
\frac{d\widehat B}{dx_f}=\kappa\int d^4\hat r\,(x_f+\xi_x)\,\xi_y^2 ,
\end{equation}
which vanishes at $x_f=0$: the rate is maximal there. Differentiating again and setting $x_f=0$, the term containing $\partial\xi_y/\partial x_f$ is multiplied by $x_f+\xi_x=0$ and drops out, leaving
\begin{equation}
\frac{d^2\widehat B}{dx_f^2}\bigg|_{x_f=0}=\kappa\int d^4\hat r\,\xi_y^2\,\psi,\qquad \psi\equiv1+\frac{\partial\xi_x}{\partial x_f}\bigg|_{x_f=0},
\end{equation}
where $\xi_y$ is the profile of the reference bounce. The function $\psi(\hat r)$ is the linear response of the inflaton to a change of its value far from the bubble. Differentiating the $\xi_x$ equation of motion gives, in the range $\widehat R\ll\hat r\ll\eta^{-1/2}$ where the mass term $\eta$ can be neglected,
\begin{equation}
\psi''+\frac{3}{\hat r}\psi'=\kappa\,\xi_y^2\,\psi,\qquad \psi'(0)=0,\qquad \psi\to1-\frac{\widehat R_c^2}{\hat r^2}\quad(\hat r\to\infty),
\label{eq:psi}
\end{equation}
which defines the length $\widehat R_c$: outside the bubble $\psi$ is a harmonic function in four dimensions, and $\widehat R_c^2$ is the coefficient of its $1/\hat r^2$ tail. The integrand above is the Laplacian of $\psi$, so the integral reduces to a surface term, $2\pi^2\hat r^3\psi'\to4\pi^2\widehat R_c^2$. With $\Delta\phi=vx_f$, $B=\widehat B/g$, and $R_c=\widehat R_c/(\sqrt gv)$,
\begin{equation}
B(\Delta\phi)\simeq B_\star+2\pi^2R_c^2(\Delta\phi)^2,\qquad c=2\pi^2\left(\frac{R_c}{R_\star}\right)^2 ,
\label{eq:cRc}
\end{equation}
which is Eq.~\eqref{eq:cmain} of the paper. The variation of the prefactor $R^{-4}$ across the window is neglected. Physically, the coupling $\kappa$ pulls the inflaton toward $\phi_\star$ inside the bubble, and the cost of nucleating at displacement $\Delta\phi$ is the gradient energy of the inflaton profile $\Delta\phi\,\psi$ that interpolates between the bubble and the background. When the coupling is strong, $\kappa\xi_y^2\hat r^2\gg1$ throughout the bubble, the inflaton is held at $\phi_\star$ inside, $R_c$ approaches the radius of that region, and $c\to2\pi^2(a/R_\star)^2$, where $a$ is the radius of that region; this equals $2\pi^2$ when $a$ coincides with the radius $R_\star$ used to normalize $c$. When it is weak, $\psi\simeq1$ and $c\simeq\kappa\int d^4\hat r\,\xi_y^2/(2\widehat R^2)$, proportional to $\kappa$. Solving Eq.~\eqref{eq:psi} on the reference bounce gives $\widehat R_c\simeq4.97$, so that $R_c/R_\star\simeq0.93$ and $c\simeq17.1$. Equation~\eqref{eq:cRc} is a local expansion about $x_f=0$; the computed action exceeds its quadratic form by less than $1\%$, or $0.03$ in $B$, across the interval $|x_f|\lesssim0.087$ that contains $99\%$ of the number integral.

\subsection{Matching and the number spectrum}
At approximately constant $H_I$, the scalar amplitude fixes the rolling speed and relates the dimensional scales to $h_\star=H_IR_\star$,
\begin{equation}
s=\frac{3\sqrt{g}h_\star^3}{2\pi\Delta_{\mathcal R}\widehat R_\star^3},\qquad v=\frac{H_I\widehat R_\star}{h_\star\sqrt{g}},\qquad \left|\frac{dN}{dx_f}\right|=\frac{3h_\star^2}{\widehat R_\star^2(s+\eta x_f)}.
\end{equation}
The reference transition is at $x_f=0$, where the longitudinal slope is $s$. The slope enters only through this mapping from $x_f$ to $N$, not through the bounce of Eq.~\eqref{eq:Ub}. The number spectrum follows by combining the nucleation rate with the dilution of bubbles formed earlier~\cite{Kleban:2023ugf,Deng:2017uwc},
\begin{equation}
\frac{1}{n_{\rm PBH}}\frac{dn_{\rm PBH}}{d\ln M}=\frac{e^{-3(N-N_\star)}}{2\mathcal P_n}\frac{\Gamma}{H_I^4},\qquad \Gamma=\frac{u}{R^4}e^{-B},\qquad \ln\frac{M}{M_\star}=2(N-N_\star).
\label{eq:benchmarkspectrum}
\end{equation}
The Jacobian $dN/d\ln M=1/2$ is included in the normalization. As in the paper, we set $u=1$ rather than calculating the fluctuation determinant~\cite{Callan:1977pt}, and impose the abundance condition
\begin{equation}
\mathcal P_n=\mathcal P_{\rm req}.
\end{equation}
Together with the amplitude relation, this fixes $g$ and $h_\star$ from the computed dimensionless profiles. For the central benchmark, $\mathcal P_{\rm req}\simeq2.2\times10^{-16}$ gives $g\simeq1.08$ and $v\simeq2.13\times10^{13}\GeV$.  The Gaussian estimate $\mathcal P=e^{-B_\star}\sqrt\pi w_N/h_\star^4$ reproduces $\mathcal P_{\rm req}$ to $1\%$ with these values. Comparing the computed width with Eq.~\eqref{eq:width} gives an effective localization coefficient $c_{\rm eff}=(2\pi\Delta_{\mathcal R}/(w_Nh_\star))^2\simeq17.5$ in the radius convention adopted here. This is about $2\%$ larger than the value $c\simeq17.1$ obtained from Eq.~\eqref{eq:cRc}, the difference arising from the terms beyond quadratic order in $B$ and the variation of the prefactor across the window, and is close to the value $c=2\pi^2\simeq19.7$ for an inflaton held fixed within radius $R$. The width $\sigma_{\ln M}\simeq0.040$ is then obtained by integrating the spectrum rather than fitting a Gaussian localization coefficient.

\subsection{Numerical checks and corrections to the frozen background}
The rate is sampled across $|x_f|\leq0.3$, covering its suppressed tails. Refining the radial grid, extending the radial domain, and widening the sampled interval in $x_f$ change the abundance and width by less than $10^{-4}$ fractionally. An independent one-field shooting calculation reproduces the central action to the same accuracy. The fluctuation check at the central bounce finds one negative radial mode and a translation mode approaching zero under refinement, as expected for the bounce~\cite{Callan:1977pt}. These checks support numerical convergence of the sampled branch, but do not exclude competing transitions.

Freezing the background neglects three effects, whose sizes we estimate. (i) Velocity. The Euclidean continuation of the slow-roll background is $x(\hat\tau)=x_f-i\dot x\,\hat\tau-\tfrac12\ddot x\,\hat\tau^2+\dots$, where overdots on $x$ denote derivatives with respect to the rescaled time $\hat t=\sqrt g\,vt$, so that $|\dot x|=|\dot\phi|/(\sqrt gv^2)\simeq7.4\times10^{-4}$ for the benchmark, where $|\dot\phi|=H_I^2/(2\pi\Delta_{\mathcal R})\simeq3.5\times10^{23}\,\GeV^2$. The velocity term is odd in $\hat\tau$, and the $O(4)$ bounce is even, so its first-order contribution to $B$ integrates to zero. The leading correction is second order in $\alpha\equiv|\dot x|\widehat R/\Delta x_{\rm loc}\simeq0.084$, where $\Delta x_{\rm loc}=w_N\,(dN/dx_f)^{-1}\simeq0.047$ is the Gaussian width of the rate in $x_f$, the interval over which $B$ rises by unity, and $dN/dx_f\simeq0.61$ for the benchmark; equivalently $\alpha=H_IR_\star/w_N$. At $x_f=0$ the inflaton induced by the velocity is $-i\dot x\,\hat\tau f_1(\hat r)$, with $f_1''+5f_1'/\hat r=\kappa\xi_y^2f_1$ and $f_1\to1$ at large $\hat r$, and including the response of the bounce gives
\begin{equation}
\delta B_{\rm vel}=-\frac{\dot x^2\kappa}{8g}\int d^4\hat r\,\hat r^2\xi_y^2f_1
=-\frac{\alpha^2}{4}\,\frac{\int d^4\hat r\,\hat r^2\xi_y^2f_1}{\widehat R^2\int d^4\hat r\,\xi_y^2\psi}\, .
\label{eq:velocitycorrection}
\end{equation}
For an inflaton held fixed within radius $\widehat R$ the ratio of integrals is $2$. Solving for $f_1$ on the reference bounce gives $2.29$, so $\delta B_{\rm vel}\simeq-0.57\,\alpha^2\simeq-4\times10^{-3}$ at the peak of the rate. (ii) Acceleration. For the slow-roll background the acceleration is suppressed relative to the slope, $|\ddot x|/s\simeq\eta/(9\widehat H^2)=\eta_V/3\simeq5\times10^{-5}$ with $\widehat H=H_I/(\sqrt gv)$, and the displacement it produces across the bounce, of order $\tfrac12|\ddot x|\widehat R^2\sim10^{-9}$, is negligible. (iii) Preparation of the state. We assume that the quantum state far from the bubble is the adiabatic vacuum of the slow-roll trajectory. In the terminology of Ref.~\cite{JKMN}, slow-roll prepares the state gradually, with no power at the transverse transition frequencies, and the transverse field is heavy, $\hat m_y/\widehat H\gg1$.  Stochastic displacements of the inflaton are suppressed relative to its classical motion by $\Delta_{\mathcal R}\simeq5\times10^{-5}$ per $e$-fold. These estimates are all well below the uncertainty in the fluctuation prefactor~\cite{Callan:1977pt,Deng:2017uwc,Kleban:2023ugf}.

\section{Applicability of the PBH mass relation}

For the local benchmark, the lower basin has vacuum density $\rho_{\rm in}\simeq V_0-\varepsilon gv^4$, with $\varepsilon gv^4/(3M_{\rm Pl}^2H_I^2)\simeq7.5\times10^{-6}$, so its interior expansion rate satisfies $H_{\rm in}\simeq H_I$. Provided that the interior remains in this basin, the bubble radius at reheating therefore obeys
\begin{equation}
R_{\rm rh}H_{\rm in}\sim e^{N_\star}\frac{H_{\rm in}}{H_I}\gg1\, .
\end{equation}
The bubbles are consequently much larger than their interior de Sitter horizon, placing them in the large supercritical regime for which the exterior black-hole mass follows the scaling adopted in Eq.~\eqref{eq:mass}~\cite{Deng:2017uwc,Kleban:2023ugf}.

\section{Acoustic normalization and localized fluctuations}
The reflecting-wall estimate of Ref.~\cite{Deng:2018cxb} gives acoustic energy $E_{\rm ac}(t)\sim GM^2/[2\lambda_D(t)]$ per PBH, where $\lambda_D$ is the physical photon-diffusion length. The dominant contribution near the thermalization epoch $t_{\rm th}$ gives
\begin{equation}
\overline\mu\sim f_i\frac{GM_i}{2\lambda_\mu}\frac{\Omega_{\rm DM}}{\Omega_r}\,, \qquad \text{with} \qquad \lambda_\mu=\frac{\lambda_D(t_{\rm th})}{a(t_{\rm th})}\, .
\label{eq:mu}
\end{equation}
The initial quantities $f_i$ and $M_i$ also describe the population during acoustic dissipation because the adopted history neglects growth before $z_i=1100$.
Using $\lambda_D^2/(Gt_{\rm th})\sim10^5M_\odot$ at $t_{\rm th}\simeq7\times10^6\,\mathrm{s}$~\cite{Deng:2018cxb} gives a comoving diffusion length $\lambda_\mu\simeq3.3\times10^{-5}\, \Mpc$. For present density parameters $\Omega_{\rm DM}$ and $\Omega_r$ of dark matter and radiation, $\Omega_{\rm DM}/\Omega_r\simeq2.9\times10^3$~\cite{Planck:2018vyg}, and substitution of Eq.~\eqref{eq:numberdensity} yields Eq.~\eqref{eq:scaledmu}. Its normalization retains the hydrodynamic uncertainty of the reflecting-wall approximation.

To estimate localized fluctuations, take a comoving Silk volume $V_S\simeq(10\Mpc)^3$~\cite{Deng:2018cxb}, with PBH count $K$, Poisson mean $\overline K=n_{\rm PBH}V_S$, and single-object contribution $\mu_1=\overline\mu/\overline K$. For comoving distance $D_{\rm LSS}\simeq14\,\mathrm{Gpc}$ to last scattering~\cite{Planck:2018vyg}, the sky contains approximately $N_S\simeq4\pi(D_{\rm LSS}/V_S^{1/3})^2$ such regions. A characteristic largest occupancy $K_{\max}$, corresponding to $\mu_{\max}=K_{\max}\mu_1$, is set by
\begin{equation}
N_S\sum_{j=K_{\max}}^\infty e^{-\overline K}\frac{\overline K^j}{j!}\sim1.
\label{eq:maximum}
\end{equation}
For the full-sample reference at $M=10^8M_\odot$ and $q=1$, this gives $K_{\max}\simeq5$ and $\mu_{\max}\sim3.2\times10^{-9}$, as quoted in the paper. The bright-subset estimate is smaller. This comparison uses the approximate rare-patch threshold discussed in Ref.~\cite{Deng:2018cxb} from Ref.~\cite{Khatri:2015tla}, which is distinct from the FIRAS monopole limit~\cite{Fixsen:1996nj}.

\section{Accretion histories and CMB response}
\label{sec:accretion_history}

\subsection{Accretion histories}
We motivate the gas supply by the limiting picture of cold, approximately spherical infall on a cosmological timescale~\cite{Ricotti:2007jk}. Equating the gravitational dynamical time $(r^3/GM)^{1/2}$ to the Hubble time $H^{-1}$ selects the radius $r=(GM/H^2)^{1/3}$, at which the characteristic gravitational speed is $(GM/r)^{1/2}=Hr$. Evaluating the spherical flux at this speed and at the mean baryon density gives
\begin{equation}
\dot M_{\rm in}=4\pi r^2\rho_bHr=\frac32\frac{\rho_b}{\rho_{\rm tot}}MH\, ,
\label{eq:accretion_supply}
\end{equation}
where $M=M(z)$ is the instantaneous mass and we have used $H^2=8\pi G\rho_{\rm tot}/3$. The order-unity normalization is an assumption, and the neglect of pressure, angular momentum, feedback, and halo structure means that Eq.~\eqref{eq:accretion_supply} should not be read as a strict bound on gas delivery. We assume that all supplied gas reaches the black hole. Accretion begins at $z_i=1100$, near recombination, when the falling free-electron abundance weakens Compton drag on the infalling gas~\cite{Ali-Haimoud:2016mbv}; earlier accretion and its delayed deposition are omitted.

For radiative efficiency $\epsilon$ the luminosity is $L=\epsilon\dot M_{\rm in}$ and the mass grows as $\dot M=(1-\epsilon)\dot M_{\rm in}$. We define the Eddington ratio $\ell\equiv L/L_E$, with $L_E=M/t_E$ and $t_E\simeq450\,\mathrm{Myr}$, and the growth factor $\mathcal G(z)\equiv M(z)/M_i$. We consider two histories. In \emph{Hubble-limited accretion} all the gas supplied by Eq.~\eqref{eq:accretion_supply} is accreted, so that, using $dt=-dz/[(1+z)H]$,
\begin{equation}
\begin{aligned}
\ell(z)&=\frac32\epsilon\frac{\rho_b}{\rho_{\rm tot}}Ht_E\, ,\\
\ln\mathcal G(z)&=\frac32(1-\epsilon)\int_z^{z_i}\frac{\rho_b(z')}{\rho_{\rm tot}(z')}\frac{dz'}{1+z'}\, .
\end{aligned}
\label{eq:accretion_growth}
\end{equation}
The Eddington ratio is independent of the mass and is largest at onset, $\ell(z_i)\simeq129\,\epsilon$. For $\epsilon=0.1$ it falls from $\ell\simeq12.9$ at $z_i$ to $5.52$ at $z=600$ and reaches unity near $z\simeq185$. In \emph{Eddington-limited accretion} the luminosity is not allowed to exceed $L_E$, which replaces the inflow by
\begin{equation}
\dot M_{\rm in}=\min\left[\frac32\frac{\rho_b}{\rho_{\rm tot}}MH,\frac{M}{\epsilon t_E}\right]\, ,
\label{eq:accretion_cap}
\end{equation}
so that $\ell\leq1$. In both histories the mass follows from
\begin{equation}
\frac{d\ln\mathcal G}{d\ln a}=\frac{\dot M}{MH}=\frac{1-\epsilon}{\epsilon}\frac{\ell}{Ht_E}\, .
\label{eq:accretion_growth_general}
\end{equation}
For $\epsilon\lesssim7.8\times10^{-3}$ the Eddington limit is never reached and the two histories coincide. In both, the inflow is proportional to $M$, so the growth factor is independent of the initial mass: accretion shifts the mass spectrum in $\ln M$ without changing $\sigma_{\ln M}$, as long as all objects share the same history. For $\epsilon=0.1$, $\mathcal G(600)\simeq1.11$ for Hubble-limited accretion, and $\mathcal G(5)\simeq2.82$ and $2.29$ for Hubble- and Eddington-limited accretion, respectively.

\subsection{Escaping radiation}
The deposition calculation follows the part of the luminosity that leaves the neighborhood of the black hole as a hard continuum with a cutoff near $200\,\mathrm{keV}$. We write it as $L_X=f_XL=\epsilon_{\rm eff}\dot M_{\rm in}$, with
\begin{equation}
\epsilon_{\rm eff}=\epsilon f_X\, ,
\label{eq:epsilon_eff}
\end{equation}
where $f_X\leq1$ is the fraction of the luminosity that both is emitted in this form and escapes. Radiation that is absorbed near the source and re-emitted below the hydrogen ionization threshold is excluded from $f_X$, while escaping ultraviolet or soft X-ray radiation would add to the ionization and strengthen the limits. Taking $f_X=1$ assigns the entire luminosity to hard escaping photons and gives the tightest limit for this spectrum. The effective efficiency $\epsilon_{\rm eff}$ controls the CMB signal, whereas $\epsilon$ alone controls the mass growth and the Eddington limit.

\subsection{Deposition and calibration}
Conservation of comoving number gives the power injected in escaping hard photons per proper volume,
\begin{equation}
\dot{\mathcal E}_{\rm inj}(z)=f_X \frac{f_i\rho_{\rm DM}(z)}{t_E}\ell(z)\mathcal G(z)\, ,
\label{eq:accretion_injection}
\end{equation}
where $\rho_{\rm DM}(z)$ is the proper dark-matter density. To account for the delay between emission and absorption, define the dimensionless deposited power
\begin{equation}
p(z)=\frac{t_E\dot{\mathcal E}_{\rm dep}(z)}{f_i\rho_{\rm DM}(z)}\, .
\label{eq:accretion_deposition}
\end{equation}
For the adopted hard continuum, the approximate Compton-loss treatment of Ref.~\cite{Ali-Haimoud:2016mbv} gives
\begin{equation}
\frac{dp}{d\ln a}=\frac{k_C}{H}f_X \ell\mathcal G-\left(4+\frac{k_C}{H}\right)p,\qquad k_C\simeq0.1n_H\sigma_T\, ,
\label{eq:accretion_transport}
\end{equation}
where $n_H$ is the total hydrogen number density, $k_C$ is the effective Compton energy-loss rate, and $\sigma_T$ is the Thomson cross section. The coefficient $4$ follows by dividing the deposited-power evolution equation, whose expansion term is $7H\dot{\mathcal E}_{\rm dep}$, by $\rho_{\rm DM}\propto a^{-3}$. We impose $p(z_i)=0$, consistent with omitting radiation emitted before onset.

We calibrate to Ref.~\cite{Serpico:2020ehh}, which uses Planck 2018, baryon acoustic oscillation, and supernova data, with ExoCLASS to compute how the deposited energy modifies the ionization history and the CMB anisotropy spectra. For PBHs radiating continuously at the Eddington luminosity it obtains the mass-independent bound $f_{\rm PBH}<2.9\times10^{-9}$. For this reference history we set $f_X=\ell=\mathcal G=1$ in Eq.~\eqref{eq:accretion_transport} and integrate from $p_{\rm Edd}(10^4)=0$, so that the solution loses memory of its initial condition before the CMB-sensitive interval.

We project the model and reference histories onto the published Planck-forecast decay-basis modes\footnote{Mode vectors and eigenvalues are in the published \href{https://faun.rc.fas.harvard.edu/epsilon/data/fits/epsilon_dec_Planck_PCs.fits}{FITS file}~\cite{Finkbeiner:2011dx}.} of Ref.~\cite{Finkbeiner:2011dx}, whose density scaling matches Eq.~\eqref{eq:accretion_deposition}. Interpolation onto the native 90-bin grid, covering approximately $10\lesssim z\lesssim1258$, gives
\begin{equation}
a_j=\sum_n e_j(z_n)p(z_n),\qquad S=\left(\sum_{j=1}^{N_{\rm PC}}\lambda_j a_j^2\right)^{1/2}\, ,
\label{eq:accretion_response}
\end{equation}
where $e_j$ and $\lambda_j$ are the mode vectors and eigenvalues, and $N_{\rm PC}$ is the number of retained principal components. The published vectors are discretely orthonormal, so no additional integration weight is introduced. Since the response is linear in the abundance,
\begin{equation}
f_{i,\max}\simeq2.9\times10^{-9}\frac{S_{\rm Edd}}{S_{\rm model}}\, .
\label{eq:accretion_calibration}
\end{equation}

The calculation uses a flat cosmology with $H_0=67.4\,\mathrm{km\,s^{-1}\,Mpc^{-1}}$, $\Omega_b=0.049$, $\Omega_m=0.315$, $\Omega_r=\Omega_m/3401$, and hydrogen mass fraction $0.755$, consistent with the \textit{Planck} reference cosmology~\cite{Planck:2018vyg}. For Hubble-limited accretion with $\epsilon=0.1$ and $f_X=1$, ten response modes give $S_{\rm model}/S_{\rm Edd}\simeq4.73$ and hence $f_{i,\max}\simeq6.13\times10^{-10}$. The first mode contains only $58\%$ of the model's squared response, but increasing from five to ten modes changes the limit only from $6.15\times10^{-10}$ to $6.13\times10^{-10}$. Eddington-limited accretion with $\epsilon=0.1$ and $f_X=1$ gives $f_{i,\max}\simeq2.86\times10^{-9}$.

The dependence on the efficiencies follows from Eq.~\eqref{eq:accretion_transport}. At fixed growth history $p$ is proportional to $f_X$, so both limits scale as $1/f_X$. For Hubble-limited accretion $\ell\propto\epsilon$, so $p\propto\epsilon f_X=\epsilon_{\rm eff}$ apart from the dependence of $\mathcal G$ on $\epsilon$, which changes the limit at the percent level for $\epsilon\lesssim0.1$. These scalings give Eq.~\eqref{eq:accretion_limits} of the paper. Limiting the inflow can only reduce the deposited power, so Hubble-limited accretion is the more constrained history at a given $\epsilon_{\rm eff}$. Photon trapping would lower $\epsilon_{\rm eff}$ and is covered by the same scaling.

\subsection{Population requirements}
Figure~\ref{fig:accretion} of the paper compares the limits of Eq.~\eqref{eq:accretion_limits} with the abundances required by Eq.~\eqref{eq:numberdensity}. The required abundance increases linearly with $M_i$, whereas the limits are independent of $M_i$ for the adopted histories. The published BHL and PR curves in that figure are shown with their fixed-mass conventions identified with the initial parameters, $M=M_i$ and $f_{\rm PBH}=f_i$. The comparison is made at accretion onset, so a black-hole mass observed at redshift $z$ is first mapped to
\begin{equation}
M_i=\frac{M(z)}{\mathcal G(z)}\, .
\label{eq:Growth}
\end{equation}
For example, an observed mass of $10^8M_\odot$ at $z=5$ corresponds to $M_i\simeq3.5\times10^7M_\odot$ for Hubble-limited and $4.4\times10^7M_\odot$ for Eddington-limited accretion with $\epsilon=0.1$. Requiring $f_{i,\rm req}\leq f_{i,\max}$ in Eq.~\eqref{eq:numberdensity} gives
\begin{equation}
q\geq q_{\min}\equiv\frac{M_i n_{\rm LRD}}{\rho_{\rm DM,0}f_{i,\max}}=\frac{M(z)n_{\rm LRD}}{\mathcal G(z)\rho_{\rm DM,0}f_{i,\max}}\, .
\label{eq:accretiontest}
\end{equation}

At $M_i=10^7M_\odot$ the bright subset requires $f_{i,\rm req}\simeq3.0\times10^{-10}/q$, so that $q_{\min}\simeq0.50\,(\epsilon_{\rm eff}/0.1)$ for Hubble-limited accretion and $q_{\min}\simeq0.10f_X$ for Eddington-limited accretion with $\epsilon=0.1$. Equivalently, the bright subset is allowed for
\begin{equation}
\frac{M_i}{M_\odot}\lesssim q
\begin{cases}
\dfrac{2.0\times10^6}{\epsilon_{\rm eff}}\,, & \text{Hubble-limited},\\[6pt]
\dfrac{9.6\times10^7}{f_X}\,, & \text{Eddington-limited, }\epsilon=0.1.
\end{cases}
\label{eq:bright_mass_limits}
\end{equation}
At the same initial mass the integrated population requires $f_{i,\rm req}\simeq(6.1$--$30)\times10^{-9}/q$, which is allowed for
\begin{equation}
\begin{aligned}
\epsilon_{\rm eff}&\lesssim\left(2\times10^{-3}\text{--}10^{-2}\right)q\left(\frac{10^7M_\odot}{M_i}\right) && \text{Hubble-limited},\\
f_X&\lesssim\left(0.1\text{--}0.5\right)q\left(\frac{10^7M_\odot}{M_i}\right) && \text{Eddington-limited, }\epsilon=0.1,
\end{aligned}
\label{eq:epsilon_limit}
\end{equation}
with the tighter value in each range corresponding to the larger number density. The X-ray fractions of observed active nuclei~\cite{Duras:2020} and the X-ray weakness of the LRDs~\cite{Yue:2024,Maiolino:2025} motivate considering $f_X<1$, although their relevance to isolated black holes before reionization is uncertain.

\subsection{Limitations}
These estimates rescale a published Planck-based constraint and are not a new likelihood analysis. They assume spatially averaged deposition, and the mean emitted power does not establish that rare sources produce a sufficiently homogeneous ionization pattern. The forecast modes adopt an ionization and heating prescription that omits the additional Lyman-$\alpha$ excitation contribution~\cite{Finkbeiner:2011dx}. The reference constraint uses spectral transfer functions in ExoCLASS~\cite{Serpico:2020ehh}, whereas both histories here are propagated with the approximate Compton-loss equation, so the calibration retains uncertainties from the response shape and the deposition treatment. A quantitative likelihood bound requires a consistent spectral deposition and ionization calculation with the relevant CMB data.

\end{document}